\documentclass[aps,prd,nofootinbib,reprint,superscriptaddress,longbibliography]{revtex4-1}
\usepackage{etoolbox}
\usepackage{dcolumn}
\usepackage{bm}
\usepackage{graphics}
\usepackage{xcolor}
\usepackage{dirtytalk}
\usepackage{graphicx} 
\usepackage{blindtext}
\usepackage{amsmath}
\usepackage{amsfonts}
\usepackage{amssymb}
\usepackage{bbold}
\definecolor{Gray}{gray}{0.85}
\definecolor{LightCyan}{rgb}{0.88,1,1}
\newcolumntype{a}{>{\columncolor{Gray}}c}
\newcolumntype{b}{>{\columncolor{white}}c}

\usepackage{mathtools,halloweenmath}
\usepackage[%
colorlinks=true,
urlcolor=blue,
linkcolor=red,
citecolor=blue
]{hyperref}
\usepackage{natbib}
\begin{document}
\title{\bf Null reduction and dynamical realization of Carrollian conformal symmetries}
\vskip 1cm
\author{Ashis Saha}
\email{ashis.saha@bose.res.in}
\affiliation{Physics and Applied Mathematics Unit,\linebreak
	Indian Statistical Institute,\linebreak
	203 B.T. Road, Kolkata 700108, India}
\affiliation{Department of Astrophysics and High Energy Physics,\linebreak
	S.N.~Bose National Centre for Basic Sciences,\linebreak
	JD Block, Sector-III, Salt Lake, Kolkata 700106, India}
\author{Rabin Banerjee}
\email{rabin@bose.res.in}
\affiliation{Department of Astrophysics and High Energy Physics,\linebreak
	S.N.~Bose National Centre for Basic Sciences,\linebreak
	JD Block, Sector-III, Salt Lake, Kolkata 700106, India}
\author{Sunandan Gangopadhyay}
\email{sunandan.gangopadhyay@bose.res.in}
\affiliation{Department of Astrophysics and High Energy Physics,\linebreak
	S.N.~Bose National Centre for Basic Sciences,\linebreak
	JD Block, Sector-III, Salt Lake, Kolkata 700106, India}	
	\begin{abstract}
\noindent We start from a Lorentzian action in a deformed light-cone background and applying the method of null reduction leads to a Carrollian action in one lower spacetime dimensions. We also identify the correct light-cone definitions of the symmetry generators and their dynamical forms in terms of the fields and take the $c\rightarrow0$ limit. It is observed that these generators produce the known kinematic Carrollian conformal algebraic commutation relations. 
\end{abstract}
\maketitle
\section{Introduction}
The investigation of symmetries of any physical theory play a crucial role in understanding the structure of that theory both at the kinematic as well as the dynamical level. In particular, study of symmetry aspects of a non-Lorentzian theory plays an important role as it provides a robust understanding of new physics which arises at both the Galilean and Carrollian limits. Mostly, this study is done at the kinematic level by making use of the In\"on\"u-Wigner contraction of the symmetry group associated to the parent theory \cite{Inonu:1953sp,Wigner:1939cj}. In this procedure, the number of generators remain the same. For instance, if we consider a conformally invariant parent theory, then one can obtain the following Carrollian conformal algebra from the In\"on\"u-Wigner contraction \cite{Chen:2023pqf}
\begin{widetext}
\begin{eqnarray}\label{CCFT}
&&\left[M^{ij},X^k\right]=i\left(\delta^{ik}X^j-\delta^{jk}X^i\right),~\left[M^{ij},M^{kl}\right]=-i\left(M_{i[} {}_l\delta_{k]}+M_{j[k}\delta_{l]i}\right),~\left[M^{i0},P^j\right]=-i\delta^{ij}P^{0}\nonumber\\
&&\left[D,H\right]=iH,~\left[D,P^i\right]=iP^i,\left[D,K^{\mu}\right]=-iK^{\mu},~\left[K^0,P^i\right]=-2iM^{i0},~\left[K^i,P^0\right]=2iM^{i0}\nonumber\\
&&\left[K^j,P^k\right]=2i\left(\delta^{jk}D+M^{jk}\right),~\left[M^{i0},K^j\right]=i\delta^{ij}K^0
\end{eqnarray}
where $P^{\mu}$ are the generators of momenta, $M^{i0}$ are boost generators, $M^{ij}$ are the generators of rotation, $D$ is dilatation generator and $K^{\mu}$ are the generators for special conformal transformations. In the above, $X=P,K,M$.
\end{widetext}
On the other hand, the light-cone formalism for quantum field theories (QFTs) provides a subtle systematic mechanism for the analysis of the symmetry properties at the dynamical level \cite{Dirac:1949cp,Brodsky:1997de,Heinzl:2000ht}. This is done by exploiting the relationship existing between the light-cone momenta and the energy-momentum stress tensor components \cite{Hagen:1972pd,Son:2008ye}. Furthermore, reduction of a null direction in the light-cone framework and taking appropriate limits reveal the underlying non-Lorentzian features of a QFT. For instance, in \cite{Julia:1994bs}, the null reduction of Einstein's theory was shown for the first time and the associated Galilean structures were explored. This in turn means that the light-cone formulation on one hand produces the non-Lorentzian versions (Galilean/Carrollian limits) of the parent theory and on the other hand it also enables one to compute the dynamical forms of the symmetry generators in terms of the fields. In this work, we shall make use of the light-cone formalism to yield a Carrollian structure \cite{SenGupta:1966qer,Duval:2014uoa,Banerjee:2024jub} of a conformally invariant QFT \cite{Bagchi:2019xfx} by using the null-reduction procedure. In this context we would like to mention that the conformal Carroll group from the perspective of Bargmann space was discussed in \cite{Duval:2014lpa}. We shall also find the correct definitions of the symmetry generators in the light-cone (and their dynamical forms) which lead us to the Carroll conformal algebraic relations given in eq.\eqref{CCFT}. The importance of our work lies in the fact that the entire analysis will be carried out from a dynamical point of view. The starting point of the analysis will be a dynamical Lorentz invariant action. Such an approach has not been taken earlier to investigate Carrollian symmetries. Some of the interesting previous studies related to Carrollian conformal symmetries can be found in \cite{Henneaux:2021yzg,Baiguera:2022lsw,Nguyen:2023vfz,Adami:2023wbe,Chen:2023pqf,Nguyen:2023miw,Bagchi:2024epw,Majumdar:2024rxg,Majumdar:2025juk,Nguyen:2025sqk}.
\section{Deformed light-cone reduction of conformally invariant field action and the Carrollian limit}
In order to proceed, let us consider the following flat metric corresponding to the background geometry (in $3+1$-spacetime dimensions)
\begin{eqnarray}
	ds^2=-c^2~dt^2+dx_1^2+dx_i^2;~i=2,3.
\end{eqnarray} 
In the light-cone coordinates, that is, $x^{\pm}=\frac{1}{\sqrt{2}}\left(ct\pm x_1\right)$, the above metric simplifies to the following form
\begin{eqnarray}
	ds^2=-2dx^+dx^-+\delta_{ij}dx^idx^j~.
\end{eqnarray}
Now, we introduce a deformation to the mentioned light-cone coordinates which is of the following form 
\begin{eqnarray}
x^+\rightarrow x^+ -\frac{\lambda}{2}x^-	
\end{eqnarray}
where $\lambda$ is a real positive number. This in turn leads us to the following deformed light-cone structure
\begin{eqnarray}
ds^2=-2dx^+dx^-+\lambda \left(dx^-\right)^2+\delta_{ij}dx^idx^j~.
\end{eqnarray}
The above form of the geometry can be realized as the flat (or trivial) pp-wave geometry \cite{Ortin2004p}. From the above line element one can note that $g_{++}=0$, $g_{--}=\lambda$, $g_{+-}=g_{-+}=-1$ and $g_{ij}=\delta_{ij}$. On the other hand, the inverse metric tensors read $g^{++}=-\lambda$, $g^{--}=0$, $g^{+-}=g^{-+}=-1$ and $g^{ij}=\delta^{ij}$. In this deformed light-cone background, we now consider the following conformally invariant action corresponding to the massless $\mathbb{C}$-scalar field theory
\begin{eqnarray}\label{orgAction}
	S=\int dx^+dx^-dx^i\left[-\frac{1}{2}g^{\mu\nu}\partial_{\mu}\phi~\partial_{\nu}\phi^{*}\right];~\phi\equiv\phi(x^{\pm},x^i)\nonumber\\
\end{eqnarray}
where $dx^i=dx^1~dx^2$. The associated energy-momentum stress tensor components read
\begin{eqnarray}\label{Stress}
	T^{\mu\nu}=-\frac{1}{2}\left(\partial^{\mu}\phi^*\partial^{\nu}\phi+\partial^{\mu}\phi~\partial^{\nu}\phi^*\right)+\frac{1}{2}g^{\mu\nu}\left(\partial_{\rho}\phi~\partial^{\rho}\phi^*\right)~.~~
\end{eqnarray}
We now write down the following components of $T^{\mu\nu}$ which will be needed in the subsequent analysis. They read
\begin{eqnarray}\label{energyM}
	T^{+-}&=&-\frac{\lambda}{2}\left(\partial_{+}\phi^*\partial_{+}\phi\right)-\frac{1}{2}\delta^{ij}\partial_{i}\phi\partial_{j}\phi^*\nonumber\\
	T^{+i}&=&\frac{\lambda}{2}\Big(\partial_{+}\phi\partial_{i}\phi^*+\partial_{+}\phi^*\partial_{i}\phi\Big)\nonumber\\
	&&+\left(\partial_{-}\phi\partial_{i}\phi^*+\partial_{-}\phi^*\partial_{i}\phi\right)\nonumber\\
	T^{++}&=&-\frac{\lambda^2}{2}\partial_{+}\phi^*\partial_{+}\phi-\partial_{-}\phi^*\partial_{-}\phi\nonumber\\
	&&-\frac{\lambda}{2}\left(\partial_{-}\phi^*\partial_{+}\phi+\partial_{-}\phi\partial_{+}\phi^*+\delta^{ij}\partial_{i}\phi\partial_{j}\phi^*\right)~.
\end{eqnarray} 
We now compactify the null direction $x^{-}$ by making use of the following coordinate transformations \cite{Banerjee:2018pvs}
\begin{eqnarray}\label{reduction}
	\phi(x^+,x^-,x^i) &=& e^{-ip_-x^-} \psi(x^+,x^i)\nonumber\\
	\phi^*(x^+,x^-,x^i) &=& e^{ip_-x^-} \psi^*(x^+,x^i)
\end{eqnarray}
where $p_-=mc$ is the conjugate momentum associated to the null direction $x^{-}$. Note that in principle, we could have taken a summation over all discrete momentum modes in the above equation. However, as we shall see in the subsequent analysis, this would only lead to an overall volume factor for the momentum modes as the exponential factors in the above equation would cancel out in the action since they appear as complex conjugates of each other.
To see this explicitly, we proceed by considering all the momentum modes for the null reduction, that is, 
	\begin{eqnarray}\label{redux2}
		\phi(x^+,x^-,x^i) &=& \int dp_- e^{-ip_-x^-} \psi(x^+,x^i)\nonumber\\
		\phi^*(x^+,x^-,x^i) &=& \int dp^{\prime}_- e^{ip^{\prime}_-x^-} \psi^*(x^+,x^i)~. 
	\end{eqnarray} 
	Substituting the above decomposition in the action given in eq.\eqref{orgAction}, we get the following expression
	\begin{eqnarray}
		S=&&2\pi\int dp_-\int dx^+~dx^i\left(\frac{1}{2}\right)\Big[\lambda\partial_{+}\psi \partial_{+}\psi^*\nonumber\\
		&&-ip_-\psi^*\overleftrightarrow{\partial_{+}}\psi-\delta^{ij}\partial_i\psi~\partial_j\psi^*\Big]~.
	\end{eqnarray}
	We now note that making the identifications $x^+=c\tau$ and $p_-=mc$, lead to the following form of the action
	\begin{eqnarray}
		S=&&\left[2\pi\int dm\right]\int  d\tau~dx^i\left(\frac{1}{2}\right)\Big[\lambda \partial_{\tau}\psi~ \partial_{\tau}\psi^*\nonumber\\
		&&-imc^2\left(\psi^*\overleftrightarrow{\partial_{\tau}}\psi\right)-c^2~\delta^{ij}\partial_i\psi~\partial_j\psi^*\Big]~.
	\end{eqnarray}
	The factor $\left[2\pi\int dm\right]$ is an overall factor and does not play any role in the rest of the analysis. We shall see subsequently that the same result can be obtained by considering eq.\eqref{reduction} but without the overall factor $\left[2\pi\int dm\right]$.\\	
At this point, we would also like to make a comment on the null reduction procedure vis-a-vis the Kaluza-Klein reduction. The null reduction leads to a non-Lorentzian theory and it is different from the standard Kaluza-Klein reduction in which one compactifies a non-null extra dimension to yield a Lorentzian theory in one lower spacetime dimension. Furthermore, the null reduction procedure requires certain limiting conditions for the speed of light $c$, that are, $c\rightarrow\infty$ for the Galilean case, and $c\rightarrow0$ for the case considered here which is the Carrollian scenario. This is fundamentally different from the standard procedure of Kaluza-Klein reduction \cite{Klein:1926tv,Salam:1981xd}.\\
Eq.\eqref{reduction} in turn leads us to the following form of the dimensionally reduced action (in $2+1$-spacetime dimensions)
\begin{eqnarray}\label{NLaction}
S&=&\int dx^+dx^i\frac{1}{2}\Big[\lambda \partial_+\psi~ \partial_+\psi^*-imc\left(\psi^*\overleftrightarrow{\partial_+}\psi\right)\nonumber\\
&&-\delta^{ij}\partial_i\psi~\partial_j\psi^*\Big]\nonumber\\
&=&\int  d\tau~dx^i\left(\frac{1}{2c}\right)\Big[\lambda \partial_{\tau}\psi~ \partial_{\tau}\psi^*-imc^2\left(\psi^*\overleftrightarrow{\partial_{\tau}}\psi\right)\nonumber\\
&&-c^2~\delta^{ij}\partial_i\psi~\partial_j\psi^*\Big]	
\end{eqnarray}
where in the last line we have identified $x^+=c\tau$ as the new time direction. By rescaling the definition of the action as $\tilde{S}=cS$ and considering the $c\rightarrow0$ limit, we obtain the Carrollian version of the parent theory. This reads
\begin{eqnarray}\label{CarrollAction}
S_{\mathrm{Carroll}}\equiv \lim_{c\rightarrow0}\tilde{S}=\int  d\tau~dx^i \frac{\lambda}{2}	\left(\partial_{\tau}\psi\right) \left(\partial_{\tau}\psi^*\right)~.
\end{eqnarray}
It is interesting to observe that one can also write down the action given in eq.\eqref{NLaction} in the following form
\begin{eqnarray}
	S&=&\int  d\tau~dx^i~c\Big[\frac{\lambda}{c^2} \partial_{\tau}\psi~ \partial_{\tau}\psi^*-im\left(\psi^*\overleftrightarrow{\partial_{\tau}}\psi\right)\nonumber\\
	&&-\delta^{ij}\partial_i\psi~\partial_j\psi^*\Big]~.
\end{eqnarray}
Now, if we rescale the definition of action as $\tilde{S}=\frac{S}{c}$ and consider the limit $c\rightarrow\infty$, we get the Schr\"odinger action
\begin{eqnarray}
	S_{\mathrm{Sch}}\equiv \lim_{c\rightarrow\infty}\tilde{S}&=&\int  d\tau~dx^i\Big[-im\left(\psi^*\overleftrightarrow{\partial_{\tau}}\psi\right)\nonumber\\
	&&-\delta^{ij}\partial_i\psi~\partial_j\psi^*\Big]~.
\end{eqnarray}
This implies that the deformed light-cone background can lead to both Schr\"odinger and Carrollian non-Lorentzian structures in one lower spacetime dimensions, based upon the choice of limit for $c$. Without loss of generality, we set $\lambda=1$ in the subsequent analysis.\\
One can now compute the conjugate momenta from the above action as $\pi_{\psi}=\frac{1}{2}\partial_{\tau}\psi^*$ and $\pi_{\psi^*}=\frac{1}{2}\partial_{\tau}\psi$ which satisfy the following commutation relations
\begin{eqnarray}\label{algebra}
\left[\psi(x^i),\pi_{\psi}(x^{\prime j})\right]&=&i\delta(x^i-x^{\prime j})\nonumber\\
\left[\psi^*(x^i),\pi_{\psi^*}(x^{\prime j})\right]&=&i\delta(x^i-x^{\prime j})~.
\end{eqnarray}
\section{Dynamical realization of conformal symmetry generators}
Our aim now is to have a dynamical realization of the symmetry generators (conformal generators) corresponding to the theory in the Carrollian limit. Further, we also want to obtain the correct form of the conformal generators in the deformed light-cone geometry. Our starting point is the definition existing between momenta and stress tensor components for a standard relativistic theory which reads \cite{Hagen:1972pd,Banerjee:2018pvs}
\begin{eqnarray}\label{starting}
P^{\mu}=\int dx^- dx^i~T^{+\mu}~.
\end{eqnarray}
The above relation helps us compute $P^{-}$, $P^+$ and $P^i$ in terms of the fields where the inputs ($T^{+\mu}$) have to be supplied from eq.\eqref{Stress}. Then one has to compactify the null direction $x^-$, identify the time $x^+=c\tau$ and take the limit $c\rightarrow0$ which will lead us to the dynamical form of the generators in terms of the fields $\psi$ and $\psi^*$. The correctness of the obtained form of the generators is then to be verified by computing the standard brackets existing between them.\\
To begin with, we start with the Hamiltonian $H$ which can be obtained from the light-cone momenta $P^-$ as
\begin{eqnarray}\label{Hamiltonian}
	H=-P^-=-\int dx^- dx^i~T^{+-}~.
\end{eqnarray}
The above definition has been given for a standard relativistic theory. We now substitute $T^{+-}$ from eq.\eqref{energyM}, compactify the null direction $x^-$ and take the Carroll limit $c\rightarrow0$. This leads to the following definition
\begin{eqnarray}\label{dynaH}
	\tilde{H}\equiv\lim_{c\rightarrow0} c^2~H=\int dx^i ~2\pi_{\psi}(x^i)\pi_{\psi^*}(x^i)~.
\end{eqnarray}
To check the consistency of this form, we compute the following algebra
\begin{eqnarray}
	 \left[\psi(x^j),\tilde{H}\right]&=&2\int dx^{\prime i} \left[\psi(x^j),\pi_{\psi}(x^{\prime i})\right]\pi_{\psi^*}(x^{\prime i})\nonumber\\
	&=& i\partial_{\tau}\psi(x^j)~.
\end{eqnarray}
This in turn means that the light-cone definition of the generator $H$ for a Lorentzian theory still produces correct time-translation in the Carrollian limit of the theory, that is 
\begin{eqnarray}
	H_{\mathrm{Carroll}}= H \equiv -\int dx^- dx^i~T^{+-}
\end{eqnarray}
where the one with subscript denotes the Carrollian counterpart of the generator. On the other hand, for the transverse momentum generator $P^i$, one has to proceed with the following light-cone definition in order to have the correct symplectic structure
\begin{eqnarray}\label{momentum}
	P^i_{\mathrm{Carroll}}=-P^i\equiv -\int dx^- dx^i~T^{+i}~.
\end{eqnarray}
By following the previously mentioned steps, one can check that the above definition leads us to the following dynamical form
\begin{eqnarray}\label{momentum2}
	\tilde{P}^i&\equiv&\lim_{c\rightarrow0}c~P_{\mathrm{Carroll}}^i\nonumber\\
	&=& -\int dx^j \left[\pi_{\psi^*}(x^j)\partial_i\psi^*(x^j)+\pi_{\psi}(x^j)\partial_i\psi(x^j)\right]~.
\end{eqnarray}
It is simple to show that the above form yields the following desired result (for space translations) associated to the Carrollian limit
\begin{eqnarray}
	\left[\psi(x^k),\tilde{P}^i\right]=-i\partial_i\psi(x^k)~.
\end{eqnarray}
Now we focus on the Carrollian boost $M^{i+}$ and rotations $M^{ij}$. One can start with the well-known form for the generator, 
 \begin{eqnarray}\label{BoostLC}
 M^{\mu\nu}=\int dx^- dx^i \left[x^{\mu} T^{+\nu}-x^{\nu} T^{+\mu}\right]~.
 \end{eqnarray}
 For $\mu=i,~\nu=+$, we get the standard expression of Lorentzian boost. We observe that the following definition yields the correct symplectic structure corresponding to the Carrollian boost
 \begin{eqnarray}
 	M^{i+}_{\mathrm{Carroll}}=-M^{i+}\equiv\int dx^- dx^i \left[x^{+} T^{+i}-x^{i} T^{++}\right]~.\nonumber\\
 \end{eqnarray}
 By following the same procedure mentioned earlier, in the limit $c\rightarrow0$, the above definition leads to the following dynamical form for the Carrollian boost
\begin{eqnarray}\label{boost}
 \tilde{M}^{i\tau}&=&\lim_{c\rightarrow0} c^2	M^{i+}_{\mathrm{Carroll}}= \int dx^j~x^i~2\pi_{\psi}(x^j)\pi_{\psi^*}(x^j)~.
 \end{eqnarray}
The generating form of the Carrollian boost can be obtained in the following way
\begin{eqnarray}
	\left[\psi(x^k),\tilde{M}^{i\tau}\right]&=&2\int dx^{\prime j} x^{\prime i}\left[\psi(x^k),\pi_{\psi}(x^{\prime j})\right]\pi_{\psi^*}(x^{\prime j})\nonumber\\
	&=&ix^i\partial_{\tau}\psi(x^k)~.
\end{eqnarray}
The boost generator plays a crucial role in this context as in case of standard Poincare algebra, two successive boosts are connected by a rotation (more specifically, Thomas-Wigner rotation) which can be understood from the result $\left[M^{i\tau},M^{j\tau}\right]=-iM^{ij}$ \cite{DiFrancesco:1997nk}.
\begin{widetext}
 However, the above result of the Carrollian boost yields 
\begin{eqnarray}
	\left[\tilde{M}^{i\tau},\tilde{M}^{j\tau}\right]=\left[\int dx^k~x^i~2\pi_{\psi}(x^k)\pi_{\psi^*}(x^k),\int dx^{\prime k}~x^{\prime j}~2\pi_{\psi}(x^{\prime k})\pi_{\psi^*}(x^{\prime k})\right]=0~.	
\end{eqnarray}	
\end{widetext}
This is indeed a feature of the non-Lorentzian physics as this result exists in both Galilean and Carrollian limit of a Poincare QFT which advocates for the fact that for non-Lorentzian QFTs, boosts become abelian. Another important algebra we should focus on is the boost-momentum algebra in the Carrollian limit. For a standard Lorentz invariant theory it generates the time-translation. However, it has been noted that for a Galilean theory, the Galilean boost commutes with the momentum \cite{Banerjee:2018pvs}. This particular result leads to the difference between the Galilean and the Schr\"odinger algebra, as the Schr\"odinger boost does not commute with the momentum and the concerned bracket introduces mass $M$ for the theory which is also the central extension of the corresponding algebra \cite{Son:2008ye}.
\begin{widetext}
In order to compute the Carrollian boost-momentum algebra we make use of the obtained expressions given in eq(s).\eqref{momentum2}, \eqref{boost}. This in turn yields
\begin{eqnarray}
\left[\tilde{M}^{i\tau},\tilde{P}^j\right]=\left[\int dx^k~x^i~2\pi_{\psi}(x^k)\pi_{\psi^*}(x^k),-\int dx^{\prime k} \left(\pi_{\psi^*}(x^{\prime k})\partial^{\prime}_j\psi^*(x^{\prime k})+\pi_{\psi}(x^{\prime k})\partial^{\prime}_j\psi(x^{\prime k})\right)\right]=-i\delta^{ij}\tilde{H}~.
\end{eqnarray}	
The above result suggests that the dynamical forms of the generators yield the desired result for a Carrollian theory.
\end{widetext}
On the other hand, we observe that the light cone definition of rotation for a standard Lorentzian theory still leads to the correct symplectic structure in the Carrollian limit. This in turn means
\begin{eqnarray}
	M^{ij}_{\mathrm{Carroll}}=M^{ij}\equiv\int dx^- dx^k \left[x^{j} T^{+i}-x^{i} T^{+j}\right]~.\nonumber\\
\end{eqnarray}
\begin{widetext}
The above definition yields the following dynamical form expressed in terms of the fields
\begin{eqnarray}\label{rotation}
	\tilde{M}^{ij}\equiv \lim_{c\rightarrow0} cM^{ij}_{\mathrm{Carroll}}&=&\int dx^k \Big[x^j\left(\pi_{\psi^*}(x^k)\partial_i\psi^*(x^k)+\pi_{\psi}(x^k)\partial_i\psi(x^k)\right)-x^i\left(\pi_{\psi^*}(x^k)\partial_j\psi^*(x^k)+\pi_{\psi}(x^k)\partial_j\psi(x^k)\right)\Big]~.\nonumber\\
\end{eqnarray}
The associated generating form of Carrollian rotations can be obtained in the following way
\begin{eqnarray}
	\left[\psi(x^l),\tilde{M}^{ij}\right]=-i(x^i\partial_j-x^j\partial_i)\psi(x^l)~.
\end{eqnarray}
\end{widetext}
The above result is what we expect in the Carrollian limit. Up to this stage, we have now derived all of the light-cone definitions for the Poincare generators in the Carrollian limit and have also determined their dynamical forms in terms of the fields. By making use of these computed forms, one can proceed to show that they satisfy the following algebraic relations
\begin{eqnarray}
\left[\tilde{M}^{ij},\tilde{M}^{k\tau}\right]&=& i\delta^{ki}\tilde{M}^{j\tau}-i\delta^{kj}\tilde{M}^{i\tau}\nonumber\\
\left[\tilde{M}^{ij},\tilde{P}^k\right]&=&i\delta^{ki}\tilde{P}^j-i\delta^{kj}\tilde{P}^i\nonumber\\
\left[\tilde{M}^{ij},\tilde{M}^{kl}\right]&=&-i\left(\tilde{M}_{i[} {}_l\delta_{k]}+\tilde{M}_{j[k}\delta_{l]i}\right)	
\end{eqnarray}
where we have used the dynamical forms given in eq.(s)\eqref{momentum2}, \eqref{boost} and m\eqref{rotation}.\\
Let us now move on to discuss the generators which extends from Poincare symmetry to the conformal symmetry, namely, dilatation and special conformal transformations (SCT). For a Lorentzian theory, the light-cone definition corresponding to dilatation reads
\begin{eqnarray}\label{dilatation}
	D&=&\int  dx^-~dx^j~g_{\mu\nu} x^{\nu} T^{+\mu}\nonumber\\
	&=&\int dx^-dx^j \left[x^{-}T^{++}+x^+T^{+-}-x^-T^{+-}-x^iT^{+i}\right]~.\nonumber\\
\end{eqnarray}
Proceeding with the above definition of the dilatation generator, we get the following result
\begin{eqnarray}
\left[\psi(x^k),D\right]&=&-i\Bigg[\frac{1}{c^2}\int dx^-x^-\partial_{\tau}+\frac{1}{c^2}\int dx^-x^-\partial_{\tau}\nonumber\\
&&+\frac{\tau}{c}\partial_{\tau}+\frac{x^j}{c}\partial_{j}\Bigg]\psi(x^k)~.	
\end{eqnarray}
The above result clearly shows that the correct transformation property for the field $\psi(x^k)$ has not been obtained from the definition of the dilatation generator given in eq.\eqref{dilatation}. This indicates that the form of the dilatation generator has to be modified. It turns out that the commutator of $\psi(x^k)$ with $\int dx^-dx^j\left(x^-T^{+-}-x^-T^{++}\right)$ yields
\begin{eqnarray}\label{modify}
	&&\left[\psi(x^k),\int dx^-dx^j\left(x^-T^{+-}-x^-T^{++}\right)\right]\nonumber\\
	&=& -i\left[\frac{1}{c^2}\int dx^-x^-\partial_{\tau}+\frac{1}{c^2}\int dx^-x^-\partial_{\tau}\right]~.
\end{eqnarray}
Adding eq.(s)(\ref{dilatation},\ref{modify}) now reveals that the following definition
\begin{eqnarray}\label{dilatationDynamical}
	\tilde{D}&\equiv& \lim_{c\rightarrow0} cD_{\mathrm{Carroll}}
\end{eqnarray}
with $D_{\mathrm{Carroll}}$ being given by
\begin{eqnarray}
	D_{\mathrm{Carroll}}=D+\int\left[x^-T^{+-}-x^-T^{++}\right]dx^-dx^j
\end{eqnarray}
yields the correct symplectic structure for the Carrollian dilatation generator
 \begin{eqnarray}
 \left[\psi(x^k),\tilde{D}\right]=-i\left(\tau\partial_{\tau}+x^j\partial_j\right)\psi(x^k)~.
 \end{eqnarray}
 It is to be noted that $D_{\mathrm{Carroll}}$ is not a symmetry of the parent (Lorentzian) theory. The dynamical form for the generator $\tilde{D}$ (defined in eq.\eqref{dilatationDynamical}) reads
\begin{eqnarray}
	\tilde{D}&=&-\int\Big(x^i\left\{\pi_{\psi}(x^j)\partial_i\psi(x^j)+\pi_{\psi^*}(x^j)\partial_i\psi^*(x^j)\right\}\nonumber\\
	&&+2\tau\pi_{\psi}(x^j)\pi_{\psi^*}(x^j)\Big)dx^j~.\nonumber\\
	\end{eqnarray} 
A similar kind of modified definition for the dilatation generator can also be observed in the Galilean limit \cite{Banerjee:2018pvs}.
 \begin{widetext}
  The obtained dynamical form of Carrollian dilatation (given in eq.\eqref{dilatationDynamical}) satisfies all of the following algebraic relations
 \begin{eqnarray}
 	\left[\tilde{D},\tilde{H}\right]=\Bigg[-\int\left(2\tau\pi_{\psi}(x^j)\pi_{\psi^*}(x^j)+x^i\left\{\pi_{\psi}(x^j)\partial_i\psi(x^j)+\pi_{\psi^*}(x^j)\partial_i\psi^*(x^j)\right\}\right)dx^j,\nonumber\\
 	\int dx^{\prime j} \left(2\pi_{\psi}(x^{\prime j})\pi_{\psi^*}(x^{\prime j})\right)\Bigg]=i\tilde{H}
 \end{eqnarray}
 and
  \begin{eqnarray}
 	\left[\tilde{D},\tilde{P}^j\right]=\Bigg[-\int\left(2\tau\pi_{\psi}(x^j)\pi_{\psi^*}(x^j)+x^i\left\{\pi_{\psi}(x^j)\partial_i\psi(x^j)+\pi_{\psi^*}(x^j)\partial_i\psi^*(x^j)\right\}\right)dx^j,\nonumber\\
 	-\int dx^{\prime k} \left(\pi_{\psi^*}(x^{\prime k})\partial^{\prime}_j\psi^*(x^{\prime k})+\pi_{\psi}(x^{\prime k})\partial^{\prime}_j\psi(x^{\prime k})\right)\Bigg]=i\tilde{P}^j
 \end{eqnarray}	
 \end{widetext}
 where one has to make use of the dynamical forms given in eq.(s)\eqref{dynaH} and \eqref{momentum2}. Finally, the Lorentzian light cone definition of SCT reads
 \begin{eqnarray}
 	K^{\mu}=\int\left[2x^{\mu}x_{\nu}T^{+\nu}-x^2T^{+\mu}\right]dx^-dx^i~.
 \end{eqnarray}
Keeping in mind the above mentioned standard definitions, we observe that $\mu=+$ yields the correct result in the Carrollian limit, that is,
\begin{eqnarray}
	K^+_{\mathrm{Carroll}}&=&K^+\equiv \int\left[2x^{+}x_{\nu}T^{+\nu}-x^2T^{++}\right]dx^-dx^i~.
\end{eqnarray}
\begin{widetext}
However, for $\mu=i$, one has to work with the following modified definition 
\begin{eqnarray}
K^{i}_{\mathrm{Carroll}}&=&-K^i\equiv \int\left[x^2T^{+i}-2x^{i}x_{\nu}T^{+\nu}\right]dx^-dx^j~.
\end{eqnarray} 
The above definitions give us the following dynamical forms in the limit $c\rightarrow0$ limit
\begin{eqnarray}\label{SCT}
	\tilde{K}^{\tau}&=&\lim_{c\rightarrow0}c^2K^+_{\mathrm{Carroll}}=\int 2x^jx_j\pi_{\psi}(x^i)\pi_{\psi^*}(x^i)dx^i\\
	\tilde{K}^{i}&=&\lim_{c\rightarrow0}cK^{i}_{\mathrm{Carroll}}\label{tSCT}\nonumber\\
	&=&-\int\Big\{4\tau x^i\pi_{\psi}(x^k)\pi_{\psi^*}(x^k)+2x^i x^j\big(\pi_{\psi}(x^k)\partial_j\psi(x^k)+\pi_{\psi^*}(x^k)\partial_j\psi^*(x^k)\big)\nonumber\\
	&&-x^2\left(\pi_{\psi}(x^k)\partial_i\psi(x^k)+\pi_{\psi^*}(x^k)\partial_i\psi^*(x^k)\right)\Big\}dx^k~.\nonumber\\\label{sSCT}
\end{eqnarray}
The SCT generators play very important role in differentiating the symmetry aspects of a standard relativistic QFT with conformal symmetry from a Carrollian QFT with the mentioned symmetry.
In case of a standard CFT, the commutator between the temporal SCT and Hamiltonian leads to dilatation, to be specific, $\left[K^{\tau},H\right]=-2iD$. However, in the Carrollian limit, we get
\begin{eqnarray}
	\left[\tilde{K}^{\tau},\tilde{H}\right]=\left[\int 2x^jx_j\pi_{\psi}(x^k)\pi_{\psi^*}(x^k)dx^k,\int dx^{\prime k} ~2\pi_{\psi}(x^{\prime k})\pi_{\psi^*}(x^{\prime k})\right]=0~.
\end{eqnarray}
\end{widetext}
This implies that they commute with each other (which can be verified by making use of the dynamical forms given in eq.(s)\eqref{dynaH} and \eqref{SCT}). This in turn means that in the limit $c\rightarrow0$, the temporal SCT and Hamiltonian become compatible symmetries and the time-evolution of the system becomes scale invariant by itself. Another interesting result which we obtain in the Carrollian limit is
\begin{eqnarray}
\left[\tilde{K}^{\tau},\tilde{M}^{i\tau}\right]=0~.
\end{eqnarray}
The above result differs from the one we see in case of a standard conformal algebra which is $\left[K^{\tau},M^{i\tau}\right]=-iK^i$. This signifies that for a Carrollian theory temporal SCT and boost are compatible symmetries.The dynamical forms of the SCT generators also lead us to the following desired Carrollian algebraic relations
\begin{widetext}
\begin{eqnarray}
&&\left[\tilde{K}^i,\tilde{P}^j\right]=2i\left(\delta^{ij}\tilde{D}+\tilde{M}^{ij}\right),~\left[\tilde{K}^i,\tilde{H}\right]=2i\tilde{M}^{i\tau},~\left[\tilde{K}^{\tau},\tilde{P}^i\right]=-2i\tilde{M}^{i\tau}\nonumber\\
&&\left[\tilde{K}^{i},\tilde{M}^{j\tau}\right]=-i\delta^{ij}\tilde{K}^{\tau},~\left[\tilde{M}^{ij}, \tilde{K}^m\right] = i\left(\delta^{im}\tilde{K}^j-\delta^{jm}\tilde{K}^i\right),~\left[\tilde{D},\tilde{K}^{\mu}\right]=-i\tilde{K}^{\mu}~.
\end{eqnarray}
\section{Conclusion}
We now summarize our findings. We observe that the light-cone definitions for the conformal generators $H$, $M^{ij}$ and $K^{+}$ remain unchanged even in the Carrollian limit. This implies that the symmetries generated by the Hamiltonian $H$, rotation generators $M^{ij}$, and the temporal special conformal transformation $K^{+}$, survive the compactification. However, for the rest of the generators some modification of the light-cone definitions are required which implies that the symmetries of the parent theory with unmodified generators do not lead to the Carrollian conformal symmetry.
\textcolor{blue}{\begin{table}[h]
	\renewcommand{\arraystretch}{2} 
	\begin{tabular}{||c|c||} 
		\hline
		Generator &  Dynamical forms \\
		\hline\hline 
		Hamiltonian &  $\tilde{H}\equiv\lim_{c\rightarrow0} c^2~H_{\mathrm{Carroll}}=\int dx^i ~2\pi_{\psi}(x^i)\pi_{\psi^*}(x^i)$\\ 
		\hline
		Momentum &  $\tilde{P}^i \equiv \lim_{c\rightarrow0}c~P_{\mathrm{Carroll}}^i= -\int dx^j \left[\pi_{\psi^*}(x^j)\partial_i\psi^*(x^j)+\pi_{\psi}(x^j)\partial_i\psi(x^j)\right]$\\
		\hline
		Boost &  $\tilde{M}^{i\tau}\equiv \lim_{c\rightarrow0} c^2	M^{i+}_{\mathrm{Carroll}}= \int dx^j~x^i~2\pi_{\psi}(x^j)\pi_{\psi^*}(x^j)$\\
		\hline
		Rotation & $\tilde{M}^{ij}\equiv \lim_{c\rightarrow0} cM^{ij}_{\mathrm{Carroll}}$~~~~~~~~~~~~~~~~~~~~~~~~~~~~~~~~~~~~~~~~~~~~~~~~~~~~~~~~~~~~~~~~~~~~~~~~~~~~~~~~~~~~~~~~~~~~~~~~~~~~~~\\
		 & $=\int dx^k \Big[x^j\left(\pi_{\psi^*}(x^k)\partial_i\psi^*(x^k)+\pi_{\psi}(x^k)\partial_i\psi(x^k)\right)-x^i\left(\pi_{\psi^*}(x^k)\partial_j\psi^*(x^k)+\pi_{\psi}(x^k)\partial_j\psi(x^k)\right)\Big]$\\
		\hline
		Dilatation &  $	\tilde{D}\equiv \lim_{c\rightarrow0} cD_{\mathrm{Carroll}}=-\int\Big(x^i\left\{\pi_{\psi}(x^j)\partial_i\psi(x^j)+\pi_{\psi^*}(x^j)\partial_i\psi^*(x^j)\right\}+2\tau\pi_{\psi}(x^j)\pi_{\psi^*}(x^j)\Big)dx^j$\\
		\hline
		SCT (spatial) &  $\tilde{K}^{i}\equiv \lim_{c\rightarrow0}cK^{i}_{\mathrm{Carroll}}$~~~~~~~~~~~~~~~~~~~~~~~~~~~~~~~~~~~~~~~~~~~~~~~~~~~~~~~~~~~~~~~~~~~~~~~~~~~~\\
		 &  $=-\int\Big\{4\tau x^i\pi_{\psi}(x^k)\pi_{\psi^*}(x^k)+2x^i x^j\big(\pi_{\psi}(x^k)\partial_j\psi(x^k)+\pi_{\psi^*}(x^k)\partial_j\psi^*(x^k)\big)$\\
		 & $-x^2\left(\pi_{\psi}(x^k)\partial_i\psi(x^k)+\pi_{\psi^*}(x^k)\partial_i\psi^*(x^k)\right)\Big\}dx^k$\\
		\hline
		SCT (temporal) &  $\tilde{K}^{\tau}\equiv \lim_{c\rightarrow0}c^2K^+_{\mathrm{Carroll}}=\int 2x^jx_j\pi_{\psi}(x^i)\pi_{\psi^*}(x^i)dx^i$\\
		\hline
	\end{tabular}
	\caption{Dynamical forms of the  conformal generators in the Carrollian limit}
	\label{Table2}
\end{table}}
\end{widetext}
In Table \eqref{Table2}, we provide the dynamical forms of the light-cone definition of the conformal generators in the Carrollian limit (that is, $c\rightarrow0$ limit) which produces the expected algebraic relations given in eq.\eqref{CCFT}. We have shown that one can obtain a Carrollian CFT from a Lorentzian CFT action by introducing a deformed light-cone background  followed by a $c\rightarrow0$ limit. Our analysis yields a Carrollian structure at a dynamical level which was missing in the previous literature. We first considered the massless complex scalar field action in this deformed light-cone background. We then compactified the undeformed null direction and identified the new time direction. Then in the limit $c\rightarrow0$, one ends up with an action invariant under the Carrollian relativistic framework. We then moved on to obtain the correct light-cone definition of the symmetry generators in the $c\rightarrow0$ limit which produces the Carrollian conformal algebraic relations. This has been done by exploiting the dynamical structure of the generators expressed in terms of the fields. We start with the energy-momentum stress-tensor corresponding to the Lorentzian parent theory and also make use of the relation existing between this and the momenta. Once again, we compactify the undeformed null direction, identify the new time-direction, and consider the limit $c\rightarrow0$. The commutators between these obtained dynamical form of the symmetry generators yields the Carrollian conformal algebra. It is important to note that this realization of the Carrollian conformal algebra has been done entirely at a dynamical level starting from a dynamical action.
\section{Acknowledgements}
The authors would like to thank the anonymous referee for very useful and crucial comments.

\bibliographystyle{hephys.bst}
\bibliography{References.bib}
\end{document}